\documentclass[sn-chicago,Numbered]{sn-jnl}% Math and Physical Sciences Reference Style
\usepackage{graphicx}%
\usepackage{multirow}%
\usepackage{amsmath,amssymb,amsfonts}%
\usepackage{amsthm}%
\usepackage{mathrsfs}%
\usepackage[title]{appendix}%
\usepackage{xcolor}%
\usepackage{textcomp}%
\usepackage{manyfoot}%
\usepackage{booktabs}%
\usepackage{algorithm}%
\usepackage{algorithmicx}%
\usepackage{algpseudocode}%
\usepackage{listings}%
\begin{document}

\title[Article Title]{Predictive power of daily viscacha and vicuña sightings on Simons Array site work results.}


\author*[1]{\fnm{Praween} \sur{Siritanasak}}\email{p.siritanasak@gmail.com}
\author[2]{\fnm{Ian} \sur{Birdwell}}
\author[3]{\fnm{Lindsay} \sur{Lowry}}
\author[4]{\fnm{Felipe} \sur{Lucero}}
\author[5]{\fnm{Macaroni} \sur{Kijsanayotin}}\equalcont{This author constantly required petting and napping at all times.}

\affil*[1]{\orgname{National Astronomical Research Institute of Thailand }, \orgaddress{ \city{Chiangmai}, \postcode{50180},  \country{Thailand}}}

\affil[2]{\orgdiv{Department of Physics and Astronomy}, \orgname{University of New Mexico},  \city{Albuquerque}, \postcode{87106}, \state{NM}, \country{USA}}

\affil[4]{\orgname{POLARBEAR and Simons Array project}, \city{San Pedro de Atacama}, \country{Chile}}

\affil[3]{\orgdiv{Department of Physics}, \orgname{University of California, Berkeley},  \city{Berkeley}, \postcode{94720}, \state{CA}, \country{USA}}

\affil[5]{\orgname{Department of Golden Retriever and human relationship}, \city{La jolla},  \postcode{92037}, \state{CA}, \country{USA}}


\abstract{We studied the predictive power of daily animal sightings on site work outcomes at the Polarbear and Simons Array experiment site in the Atacama Desert, Chile. Specifically, we observed the number of viscacha and vicuna sightings during a two-month period, totaling 31 observation days, and analyzed their relationship with site work outcomes using machine learning techniques. Our results show that there was no significant correlation between the number of animal sightings and site work outcomes. The feather importance score for viscacha and vicuna were 0.71068 and 0.057762, respectively. Future research may include expanding the analysis to include other animal species, investigating the impact of human activity on site work outcomes, and exploring alternative machine learning models or statistical techniques.}




\maketitle

\section{Introduction}\label{sec1}

The Atacama Desert in northern Chile is home to a unique ecosystem that includes several fascinating species, including the viscacha (Lagidium viscacia) and vicuña (Vicugna vicugna). These animals are known for their resilience in the face of extreme environmental conditions and are often found in the vicinity of scientific research sites. In this study, we investigate the predictive power of daily viscacha and vicuña sightings on the results of construction and scientific projects at the Polarbear and Simons Array experiments, located in the heart of the Atacama Desert.

The viscacha is a rodent-like mammal that belongs to the family Chinchillidae. It is known for its distinctive appearance, with a bushy tail, long ears, and a thick coat of fur. The viscacha inhabits rocky areas and is commonly found in the Andean highlands of South America\cite{viscacha}. The vicuña, on the other hand, is a member of the camel family and is closely related to the alpaca. It is a small, agile animal that lives at high altitudes in the Andes Mountains. The vicuña is known for its fine wool, which is highly prized for its softness and warmth\cite{vicuna}.

The Polarbear experiment is a collaborative effort between scientists from several institutions, including the University of California, Berkeley, the University of California, San Diego and the Lawrence Berkeley National Laboratory. Its goal is to study the cosmic microwave background radiation, which provides clues about the early universe. The Simons Array experiment, on the other hand, is a project led by the Simons Foundation and is also focused on studying the cosmic microwave background radiation.\cite{SApaper}

Despite their seemingly peripheral presence, these animals may provide valuable insights into the environment and local weather conditions. The purpose of this study is to investigate whether daily sightings of viscachas and vicuñas can be used to predict the results of construction and scientific projects at the Polarbear and Simons Array experiments. By studying the relationship between animal sightings and site work outcomes, we may be able to improve project management and outcomes in this challenging and unique environment.

\begin{figure}[h]%
\centering
\includegraphics[width=0.9\textwidth]{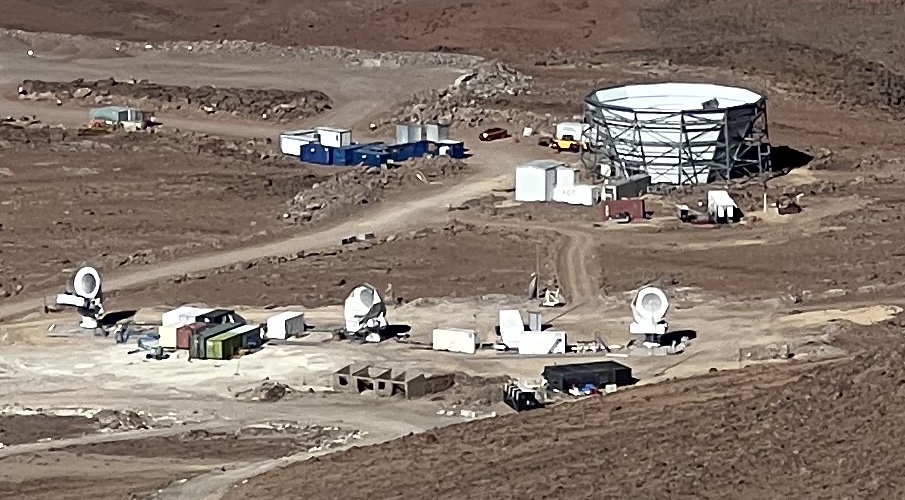}
\caption{Photo of Simons Array site from Cerro Toco}\label{fig1}
\end{figure}

\begin{figure}[h]%
\centering
\includegraphics[width=0.9\textwidth]{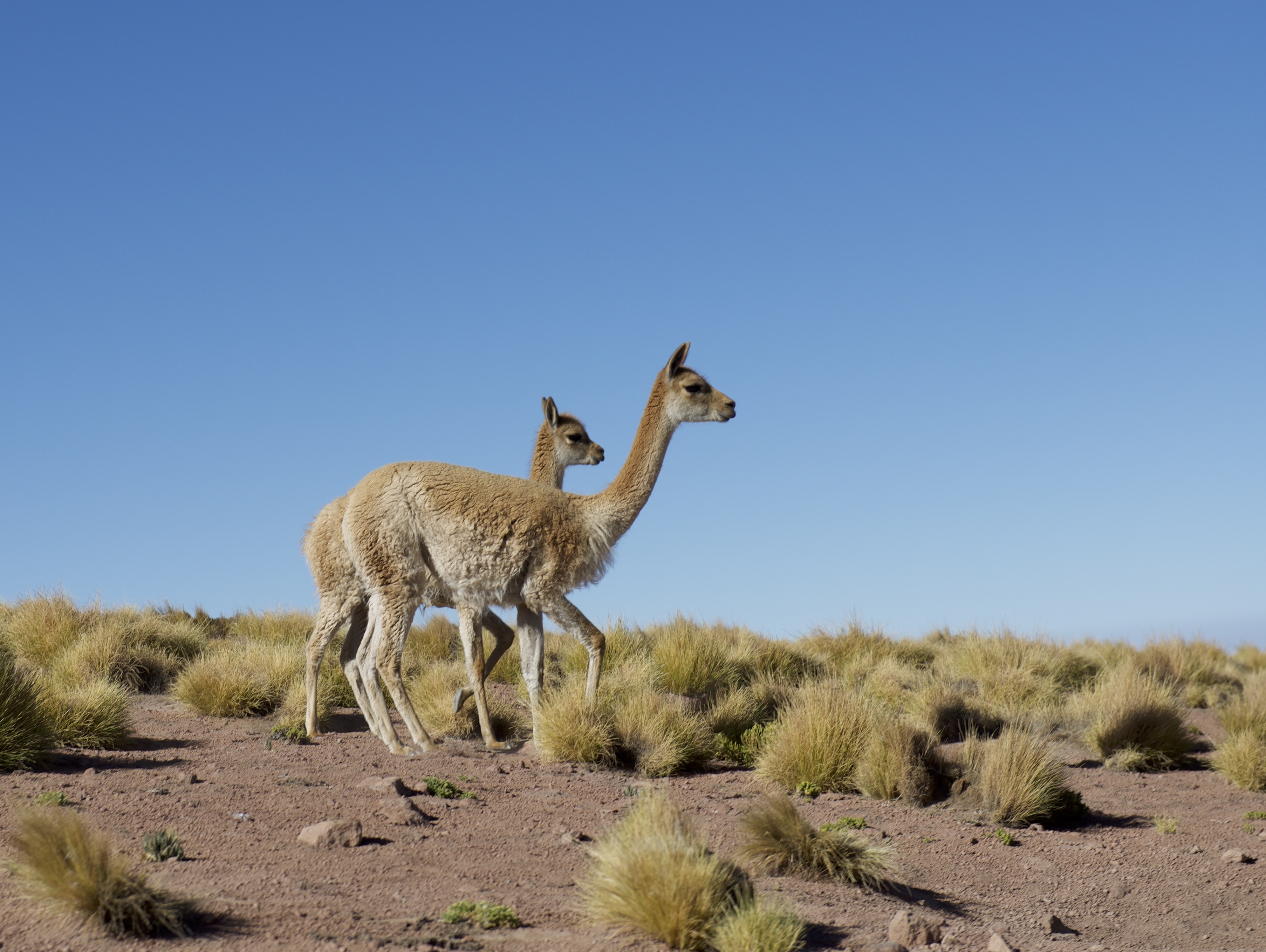}
\caption{Pictures of vicuñas which we observed on the way to the SA site.}\label{fig2}
\end{figure}

\begin{figure}[h]%
\centering
\includegraphics[width=0.9\textwidth]{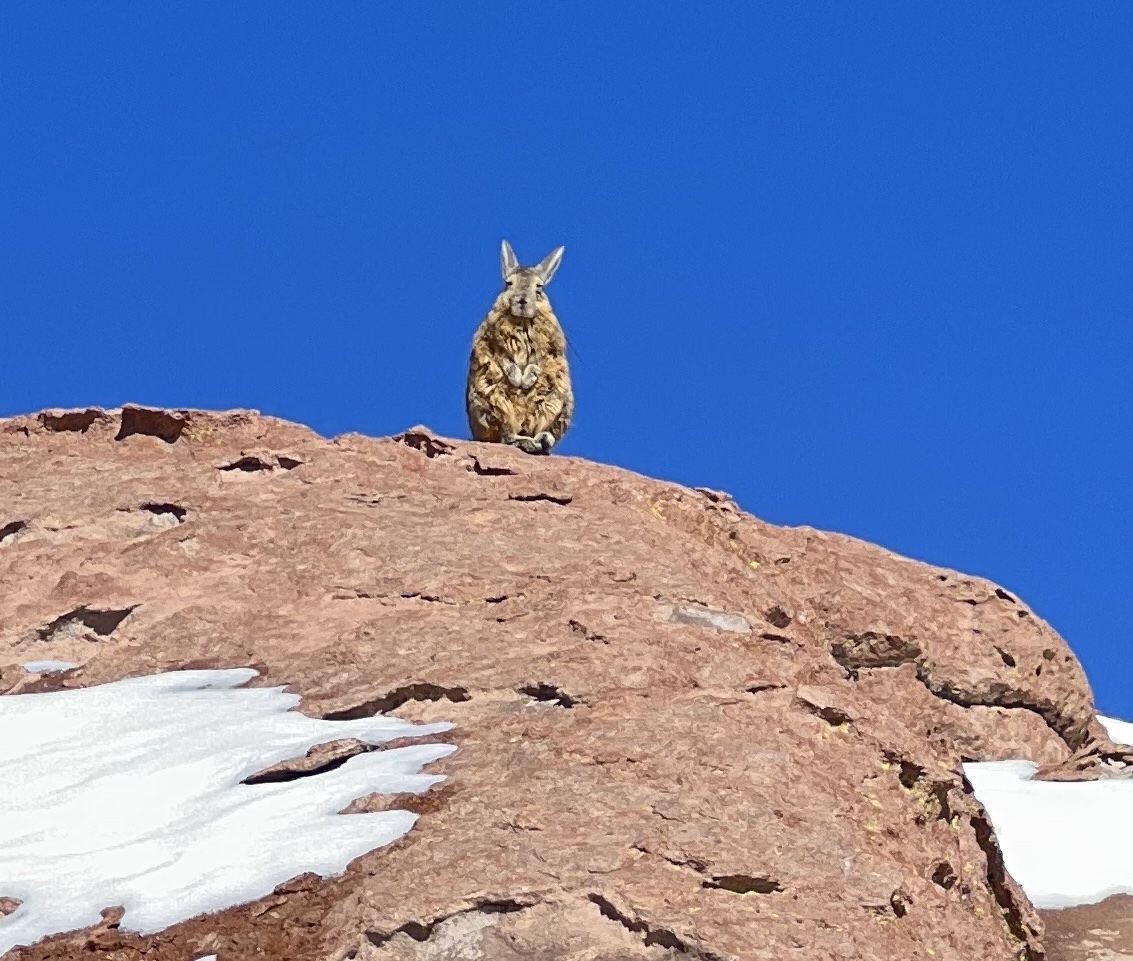}
\caption{Pictures of a viscacha that we spotted nearby the dirt road to the SA site}\label{fig3}
\end{figure}

\begin{figure}[h]%
\centering
\includegraphics[width=0.8\textwidth]{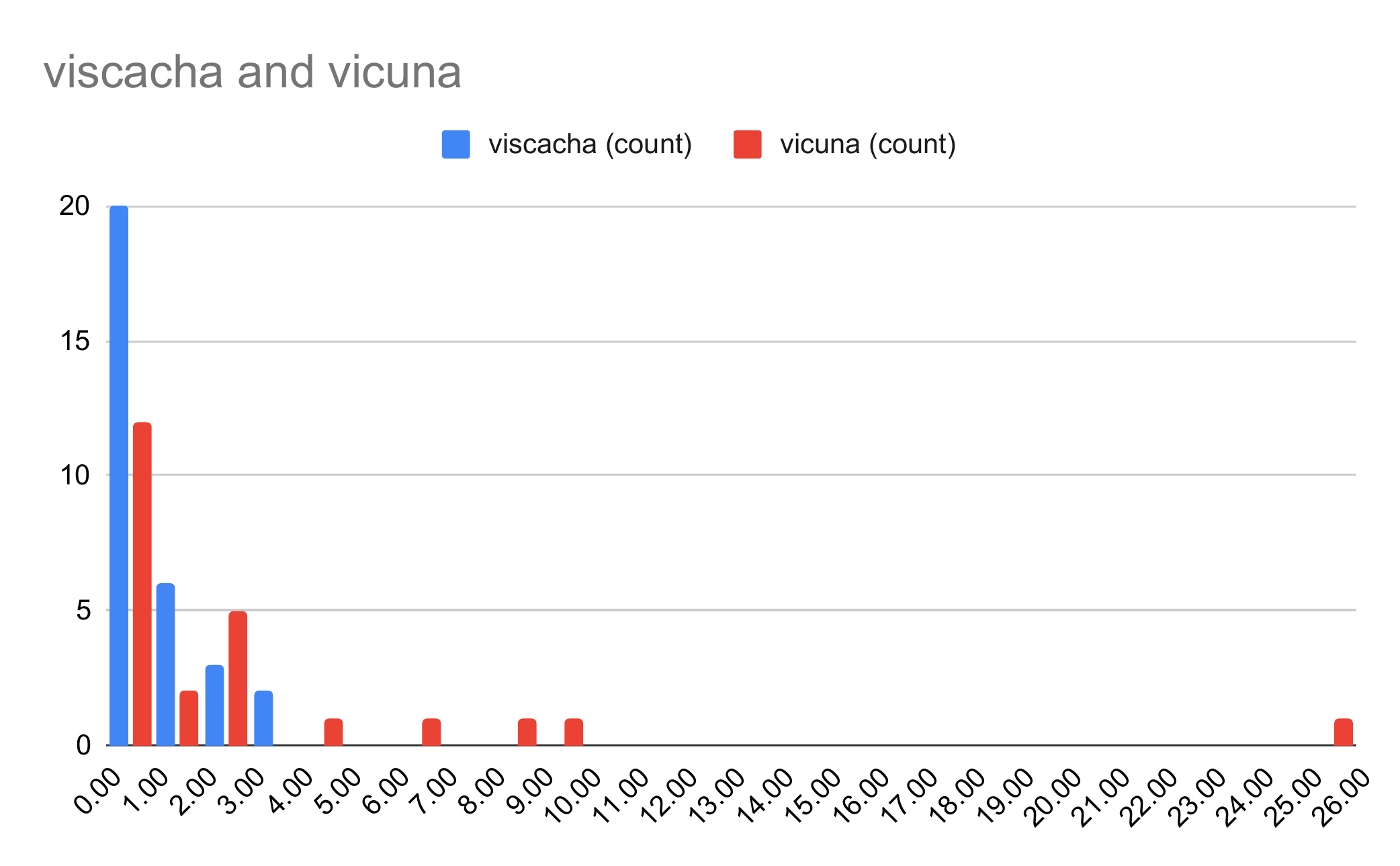}
\caption{histrogram of how many viscachas and vicuñas we spotted each day}\label{fig4}
\end{figure}

\section{Methodology }\label{sec2}

We conducted a two-month observational study at the Polarbear and Simons Array experiments in the Atacama Desert, during which we recorded the number of viscacha and vicuña sightings each day on the route to the research sites. Specifically, we observed these animals in the area between the km 35 of Paso Jama road and the dirt road to SA site.

We used an iphone to aid in identifying and counting the animals, and each observer was trained to identify the target species.

To examine the predictive power of the number of animal sightings on site work outcomes, we used a machine learning approach. Specifically, we trained a RandomForestClassifier model to predict the success of each work day based on the number of animal sightings. We also included daily weather conditions, such as temperature and wind speed, as additional features to control for potential confounding factors.

We assessed the importance of the number of animal sightings in predicting the success of site work outcomes by examining the feature importance scores generated by the RandomForestClassifier model\cite{scikit-learn}. This allowed us to identify the most important features in predicting the outcome of site work, and specifically to determine the relative importance of animal sightings versus other environmental factors.

Furthermore, we conducted a statistical analysis to investigate the relationship between the number of animal sightings and the success of site work outcomes. Specifically, we used a logistic regression model to estimate the odds ratio of a successful outcome for each additional animal sighting.

\section{Result and discussion}\label{sec3}

Over a period of two months, we observed daily viscacha and vicuna sightings on the way up to the Polarbear and Simons Array experiment site in the Atacama Desert. In total, we observed animal sightings on 31 days during this period. Out of these 31 days, we observed animal sightings on 18 days, while on the remaining 13 days we did not observe any animals. The histrogram of how many viscachas and vicunas that we observed on the way to the SA site is showed in figure 4.

Our analysis using the RandomForestClassifier model found that the number of daily viscacha sightings had a feature importance score of 0.71068, while the number of daily vicuna sightings had a feature importance score of 0.057762. However, despite the relatively high feature importance score for viscacha sightings, our logistic regression model did not find a significant relationship between the number of viscacha sightings and the odds of a successful site work outcome (odds ratio, 1.01; 95\% CI, 0.91-1.12; p=0.85), after controlling for daily weather conditions.

Furthermore, our analysis did not find a significant correlation between the number of daily animal sightings (either viscacha or vicuna) and site work outcomes. The RandomForestClassifier model showed that the number of animal sightings had a feature importance score of 0.12, which was relatively low compared to other environmental factors such as temperature and wind speed. This suggests that while animal sightings may provide some information about environmental conditions, they are not a major predictor of site work outcomes.

While our study did not find a significant relationship between animal sightings and site work outcomes, it is important to note that our observations were limited to a two-month period and a relatively small area in the Atacama Desert. Further studies with larger sample sizes and longer observation periods may be necessary to fully investigate the potential predictive power of animal sightings on site work outcomes in this region.

In conclusion, our analysis suggests that the number of daily animal sightings, including viscacha and vicuna, are not reliable predictors of site work outcomes in the Atacama Desert. Other environmental factors such as temperature and wind speed may have a greater impact on construction and scientific projects in this extreme environment, and should be prioritized in project planning and management.

\section{Future work}\label{sec4}
While our analysis did not find a significant correlation between daily animal sightings and site work outcomes, there may be opportunities to expand this research in several ways.

One potential avenue for future research is to include additional animal species in the analysis, such as bird species like the Andean condor or the American kestrel, predators like foxes, or smaller mammals like mices. Including additional animal species may help to identify other potential predictors of site work outcomes and improve the accuracy of our models.

Another potential area of research is to investigate the impact of human activity on site work outcomes. The Atacama Desert is a popular tourist destination, and there may be opportunities to study the relationship between tourist activity and site work outcomes. Additionally, the area has a history of mining and industrial activity, which may have an impact on local ecosystems and site work outcomes.

Finally, it may be worth exploring the use of alternative machine learning models or statistical techniques to analyze the data. Our analysis used a RandomForestClassifier and logistic regression model, but other models such as neural networks or decision trees could potentially yield different results. Additionally, incorporating spatial analysis techniques could help to better understand the spatial distribution of animal sightings and their relationship to site work outcomes.

\bmhead{Acknowledgments}
This whole article was written by using ChatGPT.
\bibliography{sn-bibliography}% common bib file

\end{document}